# Calcium Vulnerability Scanner (CVS): A Deeper Look

Sari Sultan and Ayed Salman

**Abstract**—Traditional vulnerability scanning methods are time-consuming and indecisive, and they negatively affect network performance by generating high network traffic. In this paper, we present a novel vulnerability scanner that is time-efficient, simple, accurate, and safe. We call it a Calcium Vulnerability Scanner (CVS). Our contribution to vulnerability scanning are the following: (i) minimize its required time and network traffic: compared to current technologies, we reduced the former by an average of 79% and the latter by 99.9%, (ii) increase its accuracy: compared to current technologies, we improved this by an average of 2600%, and (iii) enable the scanner to learn from previous scans in order to reduce future scanning time and enhance accuracy: compared to current technologies, CVS reduced scanning time by an average of 97%. CVS enables a new frontier in vulnerability scanning and allow for scalable and efficient deployment of such tools in large-scale networks, containers, edge computing, and cloud computing.

## 1 INTRODUCTION

Information technology (IT) components might suffer from security vulnerabilities that could allow adversaries to exploit them for unintended purposes. The number of new vulnerabilities per year increased by 15 times from 1999 to 2019. Additionally, a myriad of security breaches targeted virtually all aspects of the current technological landscape: afflicted individuals (e.g., identify theft), various organizations (e.g., ransomware that held many hospitals to ransoms [1]), governments (e.g., the Stuxnet worm that stymied Iranian nuclear facilities [2]), and even nations (e.g., Facebook Cambridge Analytica files that affected millions of Americans). This highlighted the importance of rigorous security testing to identify and fix vulnerabilities. Vulnerability scanning is an integral component in security assessments and risk analysis. It plays a pivotal role in safeguarding IT assets by identifying their security weaknesses, to enable fixing them before being exploited by adversaries. Due to the importance of vulnerability scanning, some governmental and industrial standards consider periodical scans as a compulsory requirement (e.g., PCI-DSS, GDPR); others consider it as a recommended requirement (e.g., ISO/IEC 27001) [3].

Vulnerability scanners evaluate targets based on a predefined policy. Policies differ in scanning times that range from several minutes to several hours per host. For example, an exhaustive scanning policy, which provides the highest level of assurance, tries to discover open ports for each connected device. Subsequently, discovered open ports are tested to identify running services. Afterwards, the scanner tries to discover vulnerabilities associated with each service. This is a protracted process that suffers from numerous issues such as (1) generating high network traffic that might congest or even disrupt network operations, (2) indecisiveness because other network devices (e.g., a firewall) might drop scanning packets without being detected, and (3) different scanning software providing different results due to lack of standardized procedures. Moreover, vulnerability scanners might alter the tested services because the scanner could send active probes that can change essential components such as databases. Sending malformed packets during scans might also contribute to destabilizing or causing a complete crash of the system [4]. Although vulnerability scanning is an important tool to assess the security of IT devices, current solutions fail to provide an easy, fast, reliable, decisive, and simple method to perform scans.

Vulnerability scanning is an essential service provided by CERTs/CSIRTs in their proactive assessments landscape. The aforementioned downsides of vulnerability scanners limit their effectiveness and scalability. Scanning a single target requires about 5 minutes to 4 hours[1]. Additionally, we found that scanning the same host more than once requires the same scanning time, as the scanner cannot improve itself from previous scans of the same host.

This led us to a fundamental research question: *is it possible for a national CERT (or any organization) to provide a vulnerability scanning service for a country that has 10 million[2] hosts?* The answer to this question is: no, not with feasible consideration for time and resources. There are two main obstacles to achieving this. First, many hosts hide behind a NAT IP address and do not have an accessible network interface outside the NAT. Deploying a localized scanner for each network might resolve this issue. However, this would lead to management overhead. Second, the time required for scanning is very long. We have been optimistic in choosing

---

*Sari Sultan, Founder of Calcium Software Solutions, Toronto, Canada, email:(sarisultan@ieee.org).*
*Ayed Salman, professor at the Department of Computer Engineering, Kuwait University, P.O. Box 5969, Safat 13060, Kuwait email:(ayed.salman@ku.edu.kw)*

1. Some references show that it takes from 20 minutes to 4 hours. However, our experimental results showed that some vulnerability scanners can finish the job in 5 minutes for TCP scans.

2. Assuming 10 million is the average country's host population.

the 5 minutes[3] per scan time; others estimated it as 20+ minutes per host. However, even with such optimism, scanning 10 million hosts, each requiring ~5 minutes, is a huge task. Increasing the level of parallelism might resolve this issue, but it would probably lead to denial of service for the CERT or hosts because the generated traffic during a scan is very large. Based on our experimental results, scanning a single host requires exchanging ~150,000 packets[4] (for TCP scans only).

**Contributions**

We propose a novel vulnerability scanner called CVS. It covers all types of vulnerabilities (e.g., network, in-host), it is fast, scalable, accurate, and learns from previous scans. We compared CVS scanner to two vulnerability scanners: Tenable Nessus and Rapid7 Insight VM (Nexpose). Compared to Nessus, CVS reduced scanning time by an average of 83% for the worst-case scenarios and by 99.75% for the best-case scenarios. Compared to Nexpose, CVS reduced scanning time by an average of 30% for the worst-case scenarios and by 98.96% for the best-case scenarios. For both cases, CVS reduced the network traffic by 99.9% and improved scanning accuracy by ~2600%. In this context, worst-case scenario refers to scanning a service for the first time, and best-case scenario means scanning a service that has been scanned previously. CVS uses the Common Platform Enumeration (CPE) to the Common Vulnerabilities and Exposure (CVE) matching to perform vulnerability scanning based on automatically generated CPEs from the properties of IT components (e.g., installed program properties in a Windows operating system). The client source code is available at https://github.com/SariSultan/CVS-client.

**Organization**

The rest of this paper is organized as follows. Section 2 supplies the background material and discusses the related work. In Section 3, we present the threat model, assumptions, and goals. Section 4 discusses our methodology and implementation. Section 5 addresses methods and conventions used to generate search components in CVS. Our experimental results are discussed in Section 6. Section 7 presents the security analysis of CVS. Finally, Section 8 concludes the paper.

## 2 BACKGROUND

The National Institute of Standards and Technology (NIST) defines a vulnerability as in Definition 1. It also defines vulnerability assessment (also known as vulnerability analysis or vulnerability scanning) as in Definition 2. Vulnerability scanning does not improve the system's security unless the discovered vulnerabilities are fixed according to the organization's remediation strategy. For example, a vulnerability scanner could discover a critical vulnerability in a device running the Windows 10 operating system. The system administrator should install the necessary patch to resolve this issue, if applicable, or restrict access to the device. Otherwise, the system remains vulnerable to exploits targeting that vulnerability.

*Definition 1 (Vulnerability).* *"weakness in an information system, system security procedures, internal controls, or implementation that could be exploited or triggered by a threat source."* [5]

*Definition 2 (Vulnerability Assessment).* *"systematic examination of an information system or product to determine the adequacy of security measures, identify security deficiencies, provide data from which to predict the effectiveness of proposed security measures, and confirm the adequacy of such measures after implementation"* [5]

To the best of our knowledge, there is no standardized way to perform vulnerability scanning [4]. However, a generalized approach followed by vulnerability scanners is shown in Algorithm 1. First, a scanner performs host discovery to identify the active targets. In a networked environment, this can be performed using different techniques such as a ping or SYN scan, among other techniques. Nmap is a prominent tool for scanning that is also used by many vulnerability scanners to perform host discovery [12]. The active hosts' ports will be scanned as well, in order to check whether the port is open or closed. Second, if the port is open, then this port scan will proceed to a service discovery process to identify the running services. Afterwards, the scanner will try different vulnerability payloads for the identified service(s) in order to check whether they are vulnerable or not. Figure 1 shows a general process for vulnerability scanning. This is a lengthy process that can take days or even weeks for scanning large networks [13].

---

**Algorithm 1:** Generalized traditional scanning process

1 **ScanTargets** ($\mathcal{T}$)
   **inputs:** $\mathcal{T}$: A set of targets to be scanned.
   **output:** $\mathcal{V}$: Vulnerabilities set (set per target $\forall t \in \mathcal{T}$). $\mathcal{V} = \{1, 2, \cdots, v, \cdots, V\}$ ($v \implies t$).
2 $\quad \mathcal{V} \leftarrow \emptyset$;
3 $\quad \mathcal{T}_{active} \leftarrow HostDiscovery(\mathcal{T})$;// Return active targets. Error $\epsilon_t$.
4 $\quad$ **foreach** $t \in \mathcal{T}_{active}$ **do**
5 $\quad\quad \mathcal{P}_{active} \leftarrow PortScan(t)$;// Return active ports for target $t$. Error $\epsilon_p$.
6 $\quad\quad$ **foreach** $p \in \mathcal{P}_{active}$ **do**
7 $\quad\quad\quad \mathcal{S}_{active} \leftarrow ServiceScan(p)$;// Return active services for port $p$. Error $\epsilon_s$.
8 $\quad\quad\quad$ **foreach** $s \in \mathcal{S}_{active}$ **do**
9 $\quad\quad\quad\quad \mathcal{V}_t +=$ $VulnerabilityScan(s)$;// Return service vulnerabilities of service $s$. Error $\epsilon_v$.
10 $\quad\quad \mathcal{V} += \mathcal{V}_t$
11 $\quad$ **return** $\mathcal{V}$;

---

3. We managed to achieve this number by disabling UDP ports scanning. Otherwise, it would take much longer to finish scans.
4. With Nessus and Rapid7 Nexpose scanners.

TABLE 1: Standards requirements related to vulnerability scanning

| Document/Standard | Description | Requirements |
| --- | --- | --- |
| PCI-DSS [6] | Standard for credit card merchants and service providers. | Requires periodical internal and external vulnerability scanning. |
| NIST 800-53 [5] | Special publication for security and privacy controls for federal information systems and organizations. | Has numerous vulnerability scanning requirements such as RA-5, CA-8, SA-11, and RA-5(4). |
| NIST Cybersecurity Framework (CSF) [7] | NIST framework for improving critical infrastructure cybersecurity. Contains best practices to improve security. | Has vulnerability scanning control DE.CM-8. |
| CIS Critical Security Controls [3] | List of 20 effective security controls for organization to improve their overall security practices. | Vulnerability scanning is ranked as one of the 20 important security controls. In particular, controls 4.1, 18.4, and 20.4 are concerned with vulnerability scanning. |
| ISO/IEC 27002:2013 [8] | Information security management system standard. This document includes the best practices for information security controls. | Vulnerability monitoring is discussed in control number 12.6.1. |
| Cloud Security Alliance (CSA) Cloud Controls Matrix (CCM) [9] | Addresses security principles important for cloud service providers. | Control TVM-02 address vulnerability scanning. |
| COBIT 4.1 & 5 [3] | Is a good-practice framework developed by ISACA for information technology (IT) management and IT governance. | Build, Acquire, and Implement (BAI) domain addresses vulnerability scanning requirements. |
| New York State Department of Financial Services 23 NY-CRR 500 [10] | Governmental standard to regulate the financial services companies. | Requires vulnerability scanning twice a year (Section 500.05). |
| Health Insurance Portability and Accountability Act (HIPAA) [11, 3] | Standards and regulations apply to health care providers. | Section 164.308(a)(ii)(A) addresses risk analysis, and Section 164.306 requires protection against possible threats. |

TABLE 2: Vulnerability scanning types

| Type | Description |
| --- | --- |
| Network-based | Performs host discovery, ports enumeration, services enumeration and tries different payloads for vulnerabilities. |
| Host-based | More comprehensive than network-based scans because it covers local and remote vulnerabilities. The target host should run the scan locally. |
| Wireless based | A special scan type to cover wireless protocols and configurations. |
| Application based | Scanning an application for application-specific vulnerabilities such as SQL injection. This is considered a risky scan because it can alter the application [3]. |

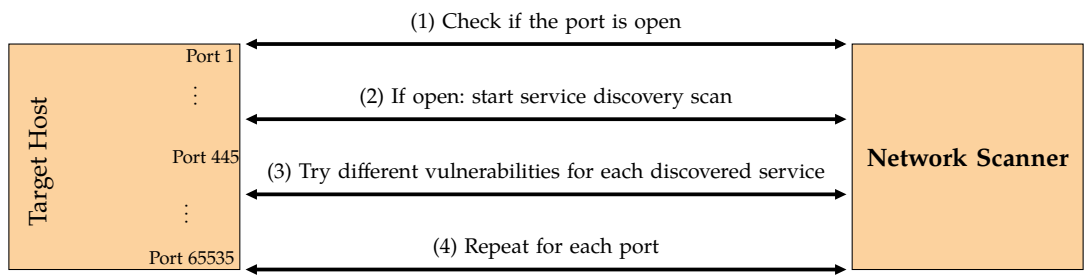

Fig. 1: Sketch of the major steps in the traditional vulnerability scanning process

Vulnerability scanning is an integral part of organizations in the 21st century because most of them rely on IT for their daily operations. Although there are no standardized methods to perform vulnerability scanning, several standards regulate vulnerability scanning outcomes. Vulnerability scanning compliance requirements are either compulsory or non-compulsory. Compulsory compliance drivers are usually governments or industry mandated requirements [3]. For example, countries could require organizations to adhere to specific rules, regulations, or standards that require performing periodical vulnerability scanning. Examples of those standards are: the Health Insur-ance Portability and Accountability Act (HIPAA), the New York State Department of Financial Services, the General Data Protection Regulation (GDPR), and the Payment Card Industry - Data Security Standard (PCI-DSS). Examples of standards that have vulnerability assessment as non-compulsory requirements are: the ISO/IEC 27001:2013, the Health Information Trust Alliance (HITRUST) CSF, and the NIST SP 800-53. Table 1 describes some standards in relation to vulnerability scanning. There are four main vulnerability scanning types: network, host, wireless, and application based [3]. Table 2 describes each of those types. In this paper, we focus on network and host-based scanning; other types are out of the scope of this tool as this moment. Although vulnerability scanning is crucial for numerous standards and organizations, academic research is limited in this area and it is dominated by the industry.

Wang and Yang [14] reviewed the major tools available for vulnerability scanning. In Table 3, we summarize important details for each of them. Although there exist numerous tools for vulnerability scanning, they all tend to share similar scanning techniques. However, different scanners produce different scanning results, so one could indicate the system is vulnerable, while another indicates that it is not. This is frustrating for users and indicates the current scanners indecisiveness. For example, El et al. [15] benchmarked numerous vulnerability scanners where results showed conspicuous differences: some scanners discovered two vulnerabilities while others discovered none. Additionally, the authors highlighted the importance of enhancing vulnerability scanning performance.

## 2.1 Internet-wide and large-scale scanning

Na et al. [16] proposed a technique for service identification based on CPEs through analyzing the service banner on open ports. We believe this work could reduce the service scanning time by targeting specific services based on CPEs' enumeration. However, accuracy could be worse than traditional scanning techniques because the authors' claim is to provide security for "Internet-connected devices" neglecting the fact that many Internet devices hide behind a NAT, which renders port scanning and service enumeration useless in most cases. Additionally, their work addresses the service discovery phase only (from phases: (i) host discovery, (ii) port discovery, (iii) service discovery, and (iv) service vulnerability scan). This means that host and port discovery phases are not enhanced. A year later, Na et al. [17] analyzed their proposed mapping technique and found that it suffers from limitations based on the CPE to vendor or product name matching. We believe their work is still in its rudimentary stages because one of their primary research goals has not been addressed, which is mapping CPEs to CVEs. The study only analyzes creating CPEs, while being oblivious to the primary goal, i.e., CVEs matching. To the best of our knowledge, this paper provides the first practical solution to map CPEs to CVEs, based on our proposed convention-based matching. Our solution is not theoretical only but has been built in practice and compared to major industrial scanners.

Some projects provide Internet-wide scanning, such as Shodan [18] and Censys [19]. These projects perform port scanning for active Internet IP addresses: if the device responds to requests, then it will be scanned and its services will be identified, based on the provided banner information. Adversaries can use this information (which is provided publicly on the project's website) to identify potential vulnerabilities in hosts. Some researchers call this operation *contact-less active reconnaissance* because the adversary does not contact the target directly [20].

Genge and Enăchescu [21] proposed a vulnerability assessment tool for Internet-connected devices, based on Shodan API. The main idea from their work is identifying CPEs based on the grabbed service version number. Na et al. [17] highlighted a limitation in the work of Genge and Enăchescu [21] which limits the match accuracy and correctness if the product name is located far from the version number. O'Hare [20] proposed a new passive vulnerability assessment tool called Scout that utilizes Censys and the National Vulnerability Database (NVD)'s feeds. Scout also tries to identify the host's vulnerabilities passively.

### 2.1.1 Critique

We believe the aforementioned studies suffer from numerous drawbacks. Firstly, almost all NAT devices are not covered by these studies and projects (unless a port forwarding feature is enabled). Secondly, many firewalls drop scanning packets; this will reduce the number of scanned hosts. Thirdly, the scanning result is not conclusive, and, as it is based on a trusted third party (i.e., Censys or Shodan), the user cannot guarantee their accuracy or freshness. Fourthly, this cannot be used for standards compliance; for example, the PCI-DSS standard requires conducting vulnerabilities assessment quarterly or after a major infrastructure upgrade; assuming that all devices within the audit scope expose their network interfaces to public IP addresses, there is still a big challenge for data freshness provided by Censys or Shodan, which might be conducted once a year, due to the large number of IPv4 addresses (i.e., ~$2^{32}$ - reserved addresses). Fifthly, assuming we are targeting IPv4, scanning the Internet means covering 4,294,967,296 total addresses - 588,514,304 reserved address = 3,706,452,992 devices, assuming each device requires 5 minutes (which is less than actual time, based on some studies [22] and our experimental results). Then, it would require the scanner ~35,748.96 years to complete the scans. Increasing the level of parallelism will decrease the time required to accomplish scanning. However, based on our experimental results, we found that current scanners require ~150,000 packets to scan a single host (for TCP scans only). Hence, increasing the level of parallelism should be associated with a rapid increase in network traffic. Even if we assume an unlimited network bandwidth, each server has a limited resources, and increasing servers would hugely increase costs. Lastly, and most importantly, not all services provide banners. There are no studies or statistics available in the literature regarding the efficacy of such solutions, but we believe accurate scanning for Internet devices is far from being realized using current technologies. CVS makes it possible.

TABLE 3: Summary of popular vulnerability assessment tools

| Tool Name | Strengths | Weaknesses |
| --- | --- | --- |
| Nessus | • Simple and easy to use<br>• Compliant with numerous standards | • Licensing fees<br>• Closed-source<br>• Scanning takes a long time |
| OpenVAS | • Compliant with numerous standards<br>• Open-source and free | • Scanning takes a long time<br>• Harder to use than alternatives |
| Nexpose | • Simple and easy to use<br>• Compliant with numerous standards | • Licensing fees<br>• Closed-source<br>• Scanning takes a long time |
| Retina CS | • Simple and easy to use<br>• Compliant with numerous standards | • Licensing fees<br>• Closed-source<br>• Scanning takes a long time |

## 3 THREAT MODEL

NIST's definition for vulnerability scanning (in Definition 2) is broad, and, based on our literature review, we believe it is hard to satisfy its requirements. Hence, we redefine vulnerability scanning as shown in Definition 3. Vulnerability scanners provide a report of vulnerabilities usually based on the CVE, the Common Vulnerability Scoring System (CVSS), CPE, and the Common Weakness Enumeration (CWE) standards. Simply put, the CVE is a list of publicly known security vulnerabilities [23]. Each CVE contains an identification number, a description, and references. CVSS is an open industry standard to characterize and score vulnerabilities [24]. CPE is a standardized method for identifying applications, operating systems, and hardware [25]. CWE is a community-developed list of software security weaknesses [26].

*Definition 3 (Vulnerability Scanning (proposed definition)).* Examining possible vulnerable components (PVCs) (see Definition 4) in order to identify vulnerabilities affecting them, based on CVE, CVSS, CPE, and CWE standards.

*Definition 4 (Possible Vulnerable Component (PVC)).* A PVC is any software or hardware that could suffer from a vulnerability. Each PVC has a required name property and optional version, vendor, edition, update, and language properties (as described in Table 4).

We consider a set of targets $\mathcal{T} = \{1, 2, \cdots, t, \cdots, T\}$, where $T = |\mathcal{T}|$ (i.e., the cardinality of set $\mathcal{T}$). We assume there exist an exhaustive set of PVCs, $\mathcal{P} = \{1, 2, \cdots, p, \cdots, P\}$, $P \in \mathbb{Z}_{>0}$. $\mathcal{P}$ contains all PVCs in NVDs database and CPE's dictionary (publicly available). Each target ($\forall t \in \mathcal{T}$) has a set of PVCs ($\mathcal{P}^t \subseteq \mathcal{P}$, where $|\mathcal{P}^t| \geq 0$). We consider a vulnerability scanning software (VSS) that has two main functionalities: (i) communicate securely with vulnerability scanning clients (VSCs) and (ii) perform the vulnerability scanning job for each VSC, where each $t \in \mathcal{T}$ has a VSC. The difference between CVS and network scanners is that the latter do not use VSCs. The VSC is a software that has one primary function: request vulnerability scan service from VSS while providing its set of PVCs ($\mathcal{P}^t$).

TABLE 4: PVC properties

| Property | Req./Opt. | Description |
| --- | --- | --- |
| Name | Required | Represents the PVC's name. Examples: Windows, Acrobat, WinRAR. |
| Version | Optional | Represents the PVC's version number. Does not have to follow any standardized naming convention. However, many vendors follow a version numbering such as {major.minor.build.revision} (e.g., 10.3.1.2344). |
| Vendor | Optional | Represents the PVC's vendor. Examples: Microsoft, Adobe, RARLabs. |
| Edition | Optional | Examples: Windows 10 Pro, Windows 7 Enterprise. |
| Update | Optional | Usually used for Windows operating systems. Examples: Service Pack 3, as in Windows XP Service Pack 3. Represents the PVC's update level for the application, it also might indicate a patch level for the PVC. |
| Language | Optional | Localization. Examples: EN-US. |

We assume the vulnerability scanning client (VSC) (i.e., CVS-client) is trusted by targets and safe. This is a reasonable assumption because the VSC is a tiny software with a single functionality, and it is open source where targets can verify that it is not collecting or reporting any personal identifiable information (PII). For more details about PII refer to [27]. We assume that the vulnerability scanning server (VSS) is honest. However, the VSC can follow honest, honest-but-curious, or malicious adversarial models. In this context, "honest" refers to respecting and not deviating from the protocol, either passively or actively. An honest-but-curious adversary is a passive adversary that will follow the protocol but will try to learn the activities of other participants. A malicious adversary is an active adversary that does not follow the protocol and tries to launch different types of attacks, such as man-in-the-middle (MITM), denial-of-service (DoS), and distributed DoS (DDoS), among other attacks. We also assume a resourceful adversary called Eve that can follow an honest-but-curious or malicious adversarial models. Eve has full access of the intermediary network between the VSCs and the VSS; she could be a system administrator or even an Internet service provider (ISP); in essence, she embodies the Dolev-Yao adversarial model [28].

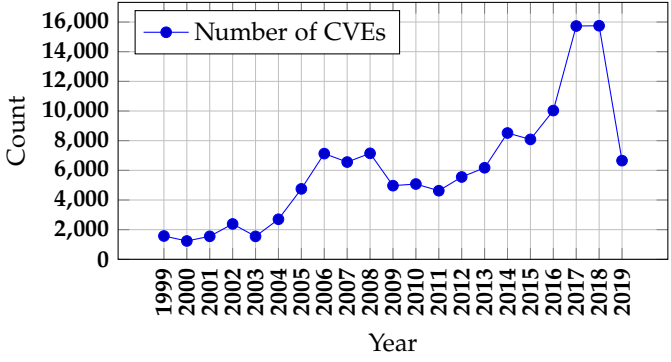

Fig. 2: NVD's database statistics for vulnerability count (as of Aug. 11, 2019)

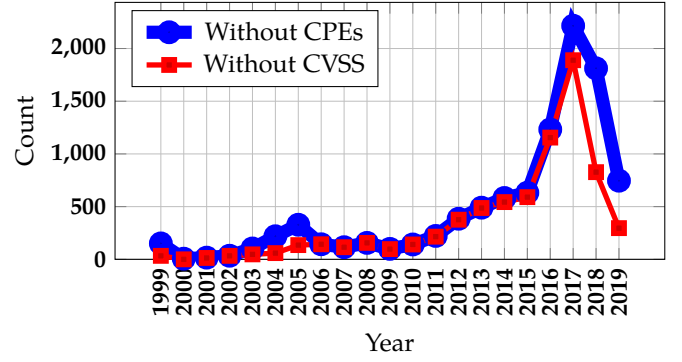

Fig. 3: NVD's database statistics for CVEs without CPE or CVSS entries (as of Aug. 11, 2019)

We assume there exist an exhaustive set of vulnerabilities, $\mathcal{V} = \{1, 2, \cdots, v, \cdots, V\}$, where $V \in \mathbb{Z}_{>0}$. There exists an exhaustive set of CVSSs, $\mathcal{CVSS} = \{1, 2, \cdots, cvss, \cdots, CVSS\}$, where $CVSS \in \mathbb{Z}_{\geq 0}$. There also exists an exhaustive set of CPEs, $\mathcal{CPE} = \{1, 2, \cdots, cpe, \cdots, CPE\}$, where $CPE \in \mathbb{Z}_{>0}$. Each vulnerability has an optional CVSS set ($\forall v \in \mathcal{V} : \exists \mathcal{CVSS}^v$, where $|\mathcal{CVSS}^v| \in \mathbb{Z}_{\geq 0}$ and $\mathcal{CVSS}^v \subseteq \mathcal{CVSS}$). Each vulnerability has an optional set of CPEs ($\forall v \in \mathcal{V} \exists \mathcal{CPE}^v$, where $|\mathcal{CPE}^v| \in \mathbb{Z}_{\geq 0}$ and $\mathcal{CPE}^v \subseteq \mathcal{CPE}$). We modeled CVSS and CPE sets as optional because our analysis showed that NVD's database contains CVEs without CVSS or CPE entries as shown in Figure 2 and Figure 3 . Definition 5 shows our formal definition for vulnerability scanning, where $\mathcal{CVE}^t$ is the set of vulnerabilities that affect the target $t$. Scan accuracy is shown in Equation (1).

***Definition 5 (Vulnerability Scanning Formal Definition (proposed definition)).***
Given $\mathcal{P}^{t*} \subseteq \mathcal{P}^t, \forall t \in \mathcal{T}$: find $\mathcal{CVE}^{t**} \subseteq \mathcal{CVE}^t$

$$\text{Scan Accuracy} = \frac{|\mathcal{CVE}^{t**} \cap \mathcal{CVE}^t|}{|\mathcal{CVE}^t|} * 100\% \quad (1)$$

Figure 4 shows the PVC's categories. Each PVC can be either vulnerable or non-vulnerable. Each vulnerable PVC can be either exploitable (if there is a known exploit for it) or non-exploitable. A non-exploitable vulnerability does not necessarily mean that it is secure because it might suffer from zero-day vulnerabilities that are not published publicly yet. Zero-day vulnerabilities are out of our scope.

In this context, our goals are the following:
- **Goal 1:** Provide a vulnerability scanning service ($\forall t \in \mathcal{T}$), based on Definition 5.
- **Goal 2:** Maximize the vulnerability scanning accuracy based on Equation (1).
- **Goal 3:** Minimize each target vulnerability scanning (i) required time, and (ii) network traffic.
- **Goal 4:** Enable the VSS to learn from previous scanning jobs (from the same target or different targets).
- **Goal 5:** The VSS and VSC, as well as the communication between them, should be protected against honest, honest-but-curious, and malicious adversaries (including Eve).

## 4 IMPLEMENTATION OVERVIEW

Figure 6 shows CVS's primary components. First, common libraries contain communication APIs and helper functions such as encryption/decryption, hashing, compression, logging, services discovery modules, and serialization. Second, the NVD contains XML schemas[5] and the classes we created to match those schemes; those classes are used as data transfer objects (DTOs). Third, the CPE's library contains the CPE XML schema and corresponding DTO classes. Fourth, the client application is a small software used by targets to request scans. We used .NET Framework 4.0 for the client application to make it support old versions of Windows OS. Currently, CVS has Windows clients only. We consider other operating systems for future work. Finally, the server module is the main part of CVS and will be discussed in detail herein. We used Entity Framework 6.0 as our object relationship mapper (ORM) and the Microsoft SQL server as our database management system (DBMS). We used C#, SQL, HTML, JavaScript (& jQuery), CSS, and XML languages to build CVS server. At the time of writing, the current CVS version is 2.0.0.0. Figure 5 shows a general overview of the scanning process in CVS.

### 4.1 CVS Server Initialization and Update

CVS is very simple and easy to use. The server application is only ~3 MB in size. We do not provide any pre-installed vulnerability sets. However, CVS should be initialized at first run, which is a common behavior among all surveyed vulnerability scanners in the literature. Figure 7 shows the server initialization process. The user starts the server by double clicking on the application executable file. Then, to initialize the server, the user should go to the configurations tab and click on "update". This will automatically download the CPEs dictionary, the NVD database, and the exploit-db sources. Each of the aforementioned will be extracted, parsed into CVS DTOs, filtered, and persisted into CVS

---
5. On Oct 16, 2019, NVD stopped providing XML feeds and moved to JSON

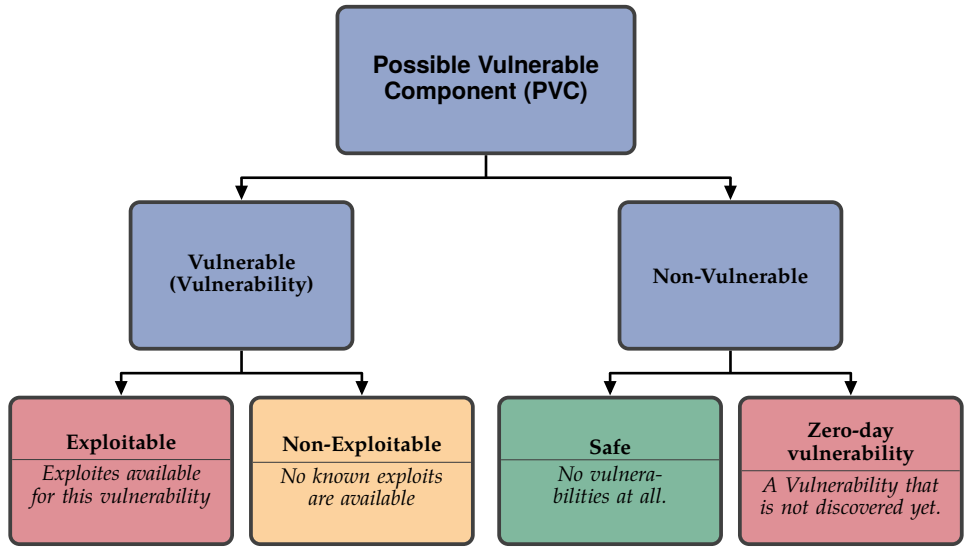

Fig. 4: Possible Vulnerable Components (PVCs) categories

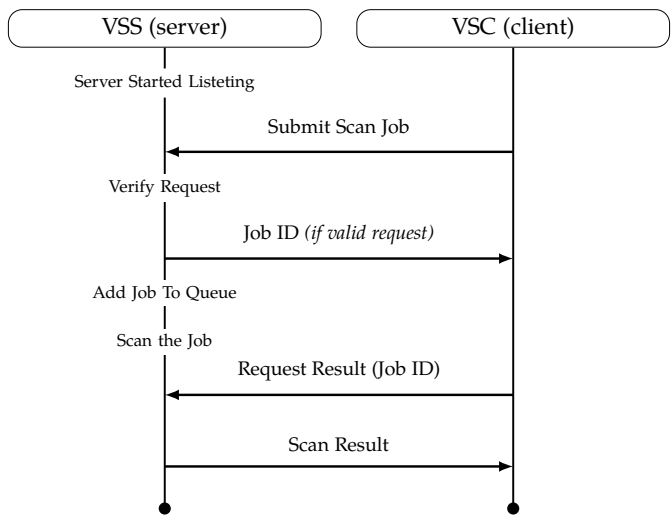

Fig. 5: Sketch of CVS scanning process

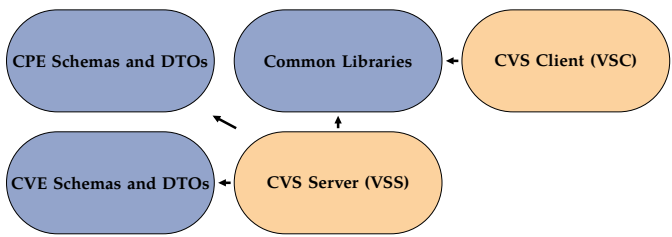

Fig. 6: CVS main components

TABLE 5: Installation requirements comparison

| Scanner | Requirements | Size | Install Time | Initialize Time |
|---|---|---|---|---|
| Nessus | CPU: Quad-core, RAM: 4 GB | 70MB | 50 S | 10m:09s |
| Nexpose | CPU: Quad-core, RAM: 8 GB | 736+MB | 1m:30s | 15m:45s |
| CVS Server | CPU: Quad-core, RAM: 8GB | 3MB | 20S | 9m:15s |

### 4.2 CVS Server Scanning Process

Figure 8 shows CVS server scanning process. It consists of five main phases: request handling, verification, queuing, scanning core service, and CPEs generation.

- **Phase 1 - Request Handling:** This phase implements the network interface and request/response operations. CVS server receives requests in a non-blocking fashion. Our initial tests show that the server can handle up to 140 concurrent requests per second (as per machine specs in Table 6). The request will be forwarded to the verification phase, which acts as a software firewall.
- **Phase 2 - Verification:** This phase acts as a rule-based software firewall. It checks the source IP/subnet, device ID, and encryption keys. It is fail-safe where the connection will be dropped if it does not comply with any of the rules.
- **Phase 3 - Queuing:** The server is built with the expectation to server a very large number of hosts concurrently. Hence, it will not be prudent to keep the requests alive until the scan job is finished. Thus, if the verification phase decided to accept the scan task, it will be added to a FIFO queue. The connection with the client will be closed and the client will be provided with a token to request the results later. The queue will serve a group of jobs concurrently, based on the hardware resources, each of which will be forwarded to the scanner phase. This implementation makes the server act somehow as a state-less server, which will help to protect against flooding attacks.

database using CVS schema. This process requires about 10 minutes (depending on Internet connection speed and workstation specs). Table 5 shows the installation requirements comparison between Nessus, Nexpose, and CVS. We believe CVS has the simplest installation and update process among surveyed scanners. Finally, the user can specify on which port to start the scanning server listening for incoming requests from VSCs i.e., CVS-clients.

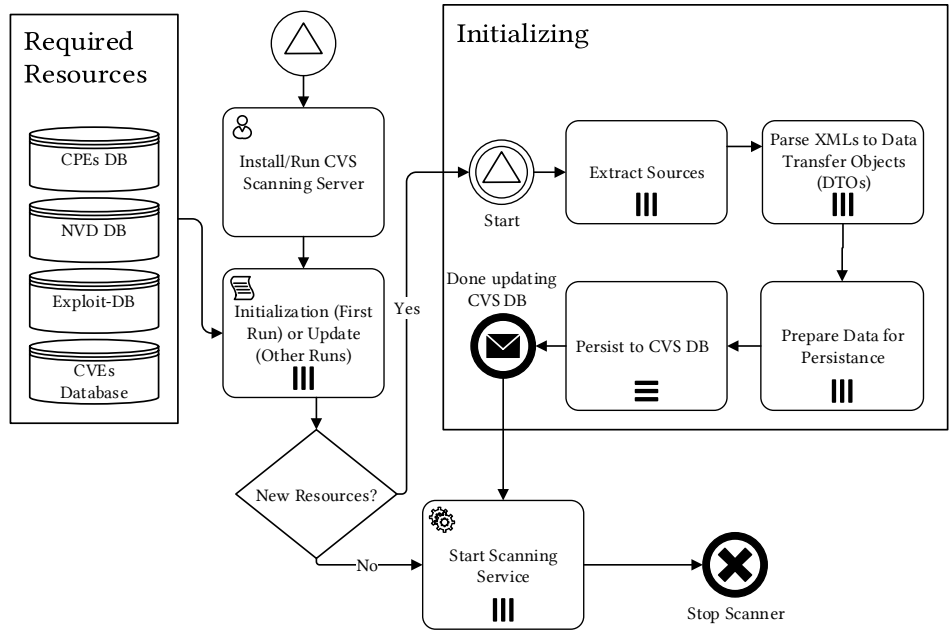

Fig. 7: CVS server initialization/update process (using BPMN notation)

TABLE 6: Scanning machines specifications

| Item | Description |
|---|---|
| For Nessus and CVS | |
| CPU | 6 cores Intel I7-8700K. |
| RAM | 8 GB. |
| OS | Windows 10 Professional. |
| Disk | SSD. |
| Network | 1Gbps Ethernet. |
| For Nexpose (because Windows 10 was not supported) | |
| CPU | 4 cores Intel I7-4700HQ. |
| RAM | 12 GB. |
| OS | Ubuntu 18.04. |
| Disk | SSD. |
| Network | 1Gbps Ethernet. |

- **Phase 4 - Scanner:** The scanner receives jobs from the queue to be scanned. For each PVC in the scan job, it will launch a thread (with a concurrency cap). Each thread will generate the possible CPEs for that PVC (as in Phase 5). Afterwards, the possible CPEs will be matched with CVS database. Then, the database will be updated in order to reduce scanning time, in case the same PVC showed up in future scans.
- **Phase 5 - CPEs Generator:** The primary function of this phase is to map PVCs to possible CPEs. The VSS uses CPEs to discover vulnerabilities targeting that PVC. This phase is discussed in detail in Section 5.

## 5 CPEs Generation

CVS supports CPEs generation for operating systems, applications, and hardware PVCs. Section 5.1 addresses generating CPEs for operating systems. Section 5.2 discusses generating CPEs for applications. Section 5.3 discusses generating CPEs for hardware. The aforementioned types share the common properties of a PVC (shown in Table 4). Hence, we made them utilize a common CPEs generation process, as shown in Algorithm 2.

### 5.1 CPEs Generation for Operating Systems

Algorithm 3 shows the CPEs generation process for operating systems. The platform CPE component is always set to 'o'. Other components are generated automatically, based on our proposed generation conventions. The CPE conventions of the operating systems are shown in Tables 12, 10, 13, 7, 8, and 9 to generate vendors, products, versions, updates, editions, and languages (respectively).

### 5.2 CPEs Generation for Applications

Algorithm 4 shows the CPEs generation process for applications. The platform CPE component is always set to 'a'. Other components are generated automatically, based on our proposed generation conventions. The application CPE conventions are shown in Tables 14, 19, 15, 16, 17, and 18 to generate vendors, products, versions, updates, editions, and languages (respectively).

### 5.3 CPEs Generation for Hardware

For hardware PVCs, we use the same process and conventions used for applications (discussed in Section 5.2). The only change is that in Algorithm 4, we change the platform from 'a' to 'h'. All previously discussed components inherit the base PVC's properties (see Table 4). We believe that the hardware PVCs will be the most challenging part for

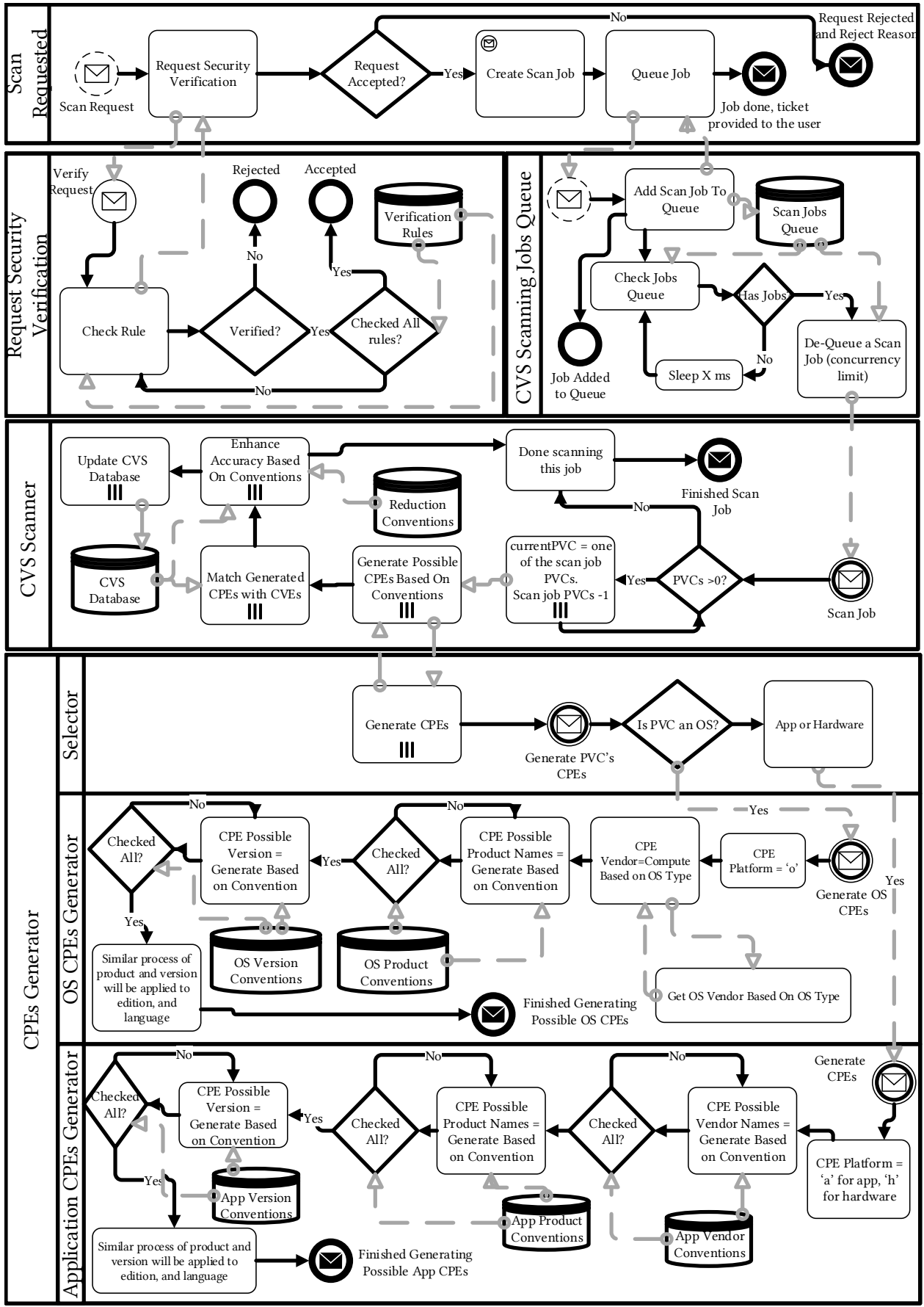

Fig. 8: CVS server scanning process (using BPMN notation)

**Algorithm 2:** CPEs generation process for PVCs

1 **PVCCPEsGeneration** ($\mathcal{P}, \mathcal{V}, \mathcal{PR}, \mathcal{VR}, \mathcal{E}, \mathcal{U}, \mathcal{L}$)
    **inputs:** $\mathcal{P}$: Set of possible platforms.
        $|\mathcal{P}| \in \mathbb{Z}_{\geq 1, \leq 3}, \forall p \in \mathcal{P} : p \notin [null, empty]$.
    **inputs:** $\mathcal{V}$: Set of possible vendor names.
        $|\mathcal{V}| \in \mathbb{Z}_{\geq 1}, \forall v \in \mathcal{V} : v \notin [null, empty]$.
    **inputs:** $\mathcal{PR}$: Set of possible product names.
        $|\mathcal{PR}| \in \mathbb{Z}_{\geq 1}, \forall pr \in \mathcal{PR} : pr \notin [null, empty]$.
    **inputs:** $\mathcal{VR}$: Set of possible versions. $|\mathcal{VR}| \in \mathbb{Z}_{\geq 1}, \forall vr \in \mathcal{VR} : vr \notin [null, empty]$.
    **inputs:** $\mathcal{E}$: Set of possible editions. $|\mathcal{E}| \in \mathbb{Z}_{\geq 0}$.
    **inputs:** $\mathcal{U}$: Set of possible updates. $|\mathcal{U}| \in \mathbb{Z}_{\geq 0}$.
    **inputs:** $\mathcal{L}$: Set of possible languages. $|\mathcal{L}| \in \mathbb{Z}_{\geq 0}$.
    **output:** $C$: Set of possible CPEs.
2   $C \leftarrow \emptyset$;
3   **foreach** ($p \in \mathcal{P}$) **do**
4     **foreach** ($v \in \mathcal{V}$) **do**
5       **foreach** ($pr \in \mathcal{PR}$) **do**
6         **foreach** ($vr \in \mathcal{VR}$) **do**
7           **if** ($\mathcal{U} = \emptyset$) **then**
8             $C$.Add("$cpe:/p:v:pr:vr$");
9           **if** ($\mathcal{E} = \emptyset$) **then**
10             $C$.Add("$cpe:/p:v:pr:vr:u$");
11           **if** ($\mathcal{L} = \emptyset$) **then**
12             $C$.Add("$cpe:/p:v:pr:vr:u:e$");
              $C$.Add("$cpe:/p:v:pr:vr:u:e:$");
13           **else**
14             **foreach** ($l \in \mathcal{L}$) **do**
15               $C$.Add("$cpe:/p:v:pr:vr:u:e:l$");
16           **else**
17             **foreach** ($e \in \mathcal{E}$) **do**
18               **foreach** ($l \in \mathcal{L}$) **do**
19                 $C$.Add("$cpe:/p:v:pr:vr:u:e:l$");
20         **else**
21           **foreach** ($u \in \mathcal{U}$) **do**
22             **foreach** ($e \in \mathcal{E}$) **do**
23               **foreach** ($l \in \mathcal{L}$) **do**
24                 $C$.Add("$cpe:/p:v:pr:vr:u:e:l$");
25   **return** $C$;

**Algorithm 3:** CPEs generation process for OSs

1 **OSCPEsGeneration** ($O$)
    **inputs:** PVC: PVC properties.
    **inputs:** $O$: OS properties. Inherits PVC properties (in Table 4).
    **output:** $C$: list of possible CPEs.
2   $C \leftarrow \emptyset$;
    /* For OS platform always assumed to be 'o' */
3   $\mathcal{P} \leftarrow$ **GenerateBasedOnConventions**('o');
    /* OS possible vendors generated according to conventions in Table 12. */
4   $\mathcal{V} \leftarrow$ **GenerateBasedOnConventions**(Table 12);
    /* OS possible product generated according to conventions in Table 10. */
5   $\mathcal{PR} \leftarrow$ **GenerateBasedOnConventions**(Table 10);
    /* OS possible versions generated according to conventions in Table 13. */
6   $\mathcal{VR} \leftarrow$ **GenerateBasedOnConventions**(Table 13);
    /* OS possible updates generated according to conventions in Table 7. */
7   $\mathcal{U} \leftarrow$ **GenerateBasedOnConventions**(Table 7);
    /* OS possible editions generated according to conventions in Table 8. */
8   $\mathcal{E} \leftarrow$ **GenerateBasedOnConventions**(Table 8);
    /* OS possible languages generated according to conventions in Table 9. */
9   $\mathcal{L} \leftarrow$ **GenerateBasedOnConventions**(Table 9);
    /* CPEs are generated based on generalized algorithm in Algorithm 2. */
10   $C \leftarrow$ **PVCCPEsGeneration**($\mathcal{P}, \mathcal{V}, \mathcal{PR}, \mathcal{VR}, \mathcal{E}, \mathcal{U}, \mathcal{L}$);
11   **return** $C$;

TABLE 7: OS update generation conventions

| # | Convention | Description | Example |
|---|---|---|---|
| 1 | Empty | As the update is an optional field | We assume the update is empty because it is optional. |
| 2 | Abbreviations | Abbreviations of service pack property | If the provided service pack is 'service pack 3' then the possible edition is 'sp3'. |

CVS deployment, because, according to our analysis, there are ~5,000 distinct vendors. It would be an intricate task to provide one client to support all hardware products because they use different libraries. On the other hand, this issue is easily resolved for operating systems and applications because they rely on a few vendors that build their software, according to general architectures (e.g., one client is enough for all Windows platforms and all applications running in Windows). One way to resolve this issue for hardware is to provide an API (e.g., RESTFUL), wherein hardware products use the API to request scans.

## 6 RESULTS AND DISCUSSION

### 6.1 Methodology

We compared CVS with two commercial vulnerability scanners: Nessus and Nexpose. We used the free trial provided by the vendors. Both scanners were configured for exhaustive scanning policy for TCP ports only. We excluded UDP

TABLE 8: OS edition generation conventions

| # | Convention | Description | Example |
|---|---|---|---|
| 1 | Empty | As the edition is an optional field | We assume the edition is empty. |

TABLE 9: OS language generation conventions

| # | Convention | Description | Example |
|---|---|---|---|
| 1 | Empty | As the language is an optional field | We assume the language is empty. |

TABLE 10: OS product name generation conventions

| # | Convention | Description | Example |
|---|---|---|---|
| 1 | Combinations | Generate combinations of the provided OS name value sub-strings based on empty spaces combined with separators in Table 11. | If the provided OS name was 'windows xp', then possible combinations will be ['windowsxp', 'windows-xp', 'windows_xp']. |

ports from scanning because this largely increases the scan

TABLE 11: Possible separator chars used for combinations

| Separator char | Description |
|---|---|
| "_" | Underscore char |
| "-" | Dash char |
| "" | none |

---

**Algorithm 4:** CPEs generation process for applications

1  **AppCPEsGeneration** ($\mathcal{A}$)
   **inputs:** PVC: PVC properties.
   **inputs:** $\mathcal{A}$: App properties. Inherits PVC properties (in Table 4).
   **output:** $C$: list of possible CPEs.
2  $C \leftarrow \emptyset$; /* App's OS platform always assumed to be 'a' */
3  $\mathcal{P} \leftarrow$ **GenerateBasedOnConventions**('a');
   /* App's possible vendors generated according to conventions in Table 14. */
4  $\mathcal{V} \leftarrow$ **GenerateBasedOnConventions**(Table 14);
   /* App's possible product generated according to conventions in Table 19. */
5  $\mathcal{PR} \leftarrow$ **GenerateBasedOnConventions**(Table 19);
   /* App's possible versions generated according to conventions in Table 15. */
6  $\mathcal{VR} \leftarrow$ **GenerateBasedOnConventions**(Table 15);
   /* App's possible updates generated according to conventions in Table 16. */
7  $\mathcal{U} \leftarrow$ **GenerateBasedOnConventions**(Table 16);
   /* App's possible editions generated according to conventions in Table 17. */
8  $\mathcal{E} \leftarrow$ **GenerateBasedOnConventions**(Table 17);
   /* App's possible languages generated according to conventions in Table 18. */
9  $\mathcal{L} \leftarrow$ **GenerateBasedOnConventions**(Table 18);
   /* CPEs are generated based on generalized algorithm in Algorithm 2. */
10 $C \leftarrow$ **PVCCPEsGeneration**($\mathcal{P}, \mathcal{V}, \mathcal{PR}, \mathcal{VR}, \mathcal{E}, \mathcal{U}, \mathcal{L}$);
11 **return** $C$;

---

time (more than an hour per host). Table 6 shows the host machines specifications. We used two machines: Windows 10 and Linux, because Nexpose did not work on the former. Table 20 shows the tested machines' specifications.

### 6.1.1 Accuracy Analysis

Recently, benchmarks have been introduced for web applications vulnerability scanners [29, 30, 31]. However, there are no benchmarks for network vulnerability scanning. Hence, to test the selected machines (in Table 20), we used a third-party statistics provider to get the actual number of vulnerabilities in each operating system. We reduced the results comparison problem for operating system vulnerabilities only, neglecting any installed applications, to make it easier to compare. Even without reducing the discovered number of vulnerabilities by Nessus and Nexpose, they were very few, ranging from 4 to 6 vulnerabilities. However, the actual vulnerabilities based on [32] were 180, 199, and 391 for Windows server 2008[6], Windows 7[7], and Windows XP [8] (respectively). Additionally, we assumed that all vulnerabilities reported by Nessus and Nexpose were correct, to show the upper limit for their accuracy. We calculated their accuracy by dividing the number of their discovered vulnerabilities by the actual vulnerabilities. For CVS, however, we first matched each of the discovered vulnerabilities with the actual vulnerabilities set, then accuracy was calculated by dividing the matches over the actual number of vulnerabilities, as in Equation (1), to show the actual accuracy of CVS.

### 6.2 Results

#### 6.2.1 Performance Results

Table 21 shows our performance analysis results and comparison between Nessus, Nexpose, and CVS. We focused on two performance measurements: scanning time and communicated network traffic. CVS had the lowest scanning time and communicated network traffic for all experiments. Compared to Nessus, CVS reduced scanning time and network traffic by an average of 93.59% and 99.97% (respectively). Compared to Nexpose, CVS reduced scanning time and network traffic by an average of 64.92% and 99.97% (respectively).

Additionally, we scanned each host two times using the same scanner to check whether the scanner learns from previous scans or not. There was no improvement for Nessus and Nexpose. However, CVS improved scanning time, with an average of ~97% when the same host was scanned for the second time, because CVS learns from previous experiences and the scanning engine can identify previous PVCs that are impossible to change unless the database is updated. For each database update, however, PVCs scanning results records will be adapted to the new updated vulnerabilities.

---

6. https://www.cvedetails.com/vulnerability-list/vendor_id-26/product_id-11366/version_id-82464/Microsoft-Windows-Server-2008--.html
7. https://www.cvedetails.com/vulnerability-list/vendor_id-26/product_id-17153/version_id-138704/Microsoft-Windows-7--.html
8. https://www.cvedetails.com/vulnerability-list/vendor_id-26/product_id-739/version_id-46866/Microsoft-Windows-Xp-.html
4

TABLE 12: OS vendor name generation conventions

| # | Convention | Description | Example |
|---|---|---|---|
| 1 | WinOS.A | For Windows OS, the vendor always set to 'microsoft'. | Windows XP, Windows 10. |
| 2 | Android.A | Match vendor with selected android possible vendors from CVS DB. Currently Android has 3 vendors (Google, Motorolla, and CodeAurora). If could not discover correct vendor, then use all vendors from possible vendors in CVS DB. | cpe:/o:**motorola**:android:4.1.2, cpe:/o:**google**:android:8.0, cpe:/o:**codeaurora**:android-msm:3.2.57 |
| 3 | Apple.A | Currently Apple has the following OSs [a_ux, airport_base_station_firmware, apple_tv, iphone_os, mac_os, mac_os_appleshare_ip, mac_os_server, mac_os_x, mac_os_x_server, os_x_server, watch_os, watchos]. Hence, match PVC properties to one of them. (The list is updated automatically when new OSs are added.) | cpe:/o:**apple**:iphone_os:10.1.1. |
| 4 | Linux.A | Currently Linux has 22 possible vendors [canonical, conectiva, corel, debian, engardelinux, gentoo, ibm, linux, linuxmint, mandrakesoft, mandriva, novell, opensuse, opensuse_project, oracle, redhat, scientificlinux, sgi, slackware, suse, trustix, windriver]. Each vendor has a set of operating system names associated with it. Hence, first try to match PVC properties to one of the vendors directly. Second, if it did not match, then try to match PVC properties to OS name, then reflect on its vendor (updated automatically from CVS DB). | Ubuntu Linux will be matched to vendor name canonical. |
| 5 | Other.A | Find a list of possible vendors from CVS DB, then try to match PVC properties with one or more of them. | - |
| 6 | Other.B | Find a list of possible vendors from DB, and for each vendor, find the list of possible OS names. Then, try to match PVC properties to the OS name; if matched, then reflect this on the matched OS name vendor. | - |

TABLE 13: OS version generation conventions

| # | Convention | Description | Example |
|---|---|---|---|
| 1 | M.M.B. | The version is major.minor.build | Create version as "Major. Minor. Build" 10.1.2991. |
| 2 | Revision | The version is equal to revision | Assume that the OS version is equal to the OS revision. |
| 3 | M.M.B.R. | major.minor.build.revision | Format as "Major.Minor.Build.Revision" 10.1.2991.5000. |
| 4 | M.M. | major.minor only | Format as "Major.Minor" 10.1. |
| 5 | Build | Build only | Use OS info properties to generate version in the format of "Build". |
| 6 | dash | Use - | Assume the version is '-'; this is sometimes used to indicate a vulnerability in all versions of the OS. |
| 7 | Combinations | OS name combinations | Similar to convention 1 from Table 10. Some OSs use the name combination for the version. |

TABLE 14: App vendor generation conventions

| # | Convention | Description | Example |
|---|---|---|---|
| 1 | App publisher property | If the publisher property is not empty, we assume the publisher is the vendor. | As 'Adobe' for 'Adobe Reader'. |
| 2 | First word | If the application name contains two or more words, then the vendor is the first word. | If app name is 'Adobe Reader', we assume vendor is 'Adobe'. |
| 3 | First and second word | If the application name contains three or more words, then the vendor is the first and second words combined with separators Table 11. | "Acme solutions paint" then we assume the vendors are ["Acmesolutions", "Acme_solutions", "Acme-solutions]. |

TABLE 15: App version generation conventions

| # | Convention | Description | Example |
|---|---|---|---|
| 1 | Display version property | If the display version property is not empty we assume it is the version. | See https://goo.gl/PJoWKM for products on Windows platform. |
| 2 | Version in title | Check the app display name if it contains version strings using Regex. | 'CPUID 1.82.2', then version is 1.82.2. |

TABLE 16: App update generation conventions

| # | Convention | Description | Example |
|---|---|---|---|
| 1 | Empty | As the update is an optional field | We assume the update is empty. |

TABLE 17: App edition generation conventions

| # | Convention | Description | Example |
|---|---|---|---|
| 1 | Empty | As the edition is an optional field | We assume the edition is empty. |

TABLE 18: App language generation conventions

| # | Convention | Description | Example |
|---|---|---|---|
| 1 | Empty | As the language is an optional field | We assume the language is empty. |

TABLE 19: App product name generation conventions

| # | Convention | Description | Example |
|---|---|---|---|
| | | Phase 1: If product exists in exhaustive products list in CVS DB, skip to Phase2 | |
| P1.1 | Combinations | Generate combinations of the provided OS name value sub-strings based on empty spaces combined with separators in Table 11, removing strings that match version format of major.minor.* . | If the provided app name was "visual studio 14.1", then possible combinations will be ["visualstudio", "visual-studio", "visual_studio"]. |
| P1.2 | Abbreviations | Abbreviations of product name, removing strings that match version format of major.minor.* . | If the provided app name was "media player classic 12.5", then possible combinations will be ["mpc"]. |
| | | Phase2: Only applied if Phase 1 failed to get match. | |
| P2.1 | First word | If the product name has only one word, we assume it is the product name. | |
| P2.2.A | Name has two words | If the product name has two non-version words, then the name is the first word. | App name "java 7 update 1", possible products are ["java"]. |
| P2.2.B | Name has two words | If the product name has two non-version words, then the name is their combinations with separators in Table 11 | App name "github desktop", possible products are ["githubdesktop", "github_desktop", "github-desktop"]. |
| P2.3.A | Name has three words | If the product name has three non-version words, then the name is the second word. | App name "jetbrains resharper ultimate", possible products are ["resharper"]. |
| P2.3.B | Name has three words | First and third word, or second and third word combined with separators in Table 11. | "vlc media player", possible products ["vlcmediaplayer", "vlc_media_player", "vlc-media-player"]. |
| P2.4.B | Name has more than three words | All words combined with separators in Table 11. | |

TABLE 20: Tested machines specifications

| Machine | Description |
|---|---|
| Windows XP SP3 build 5.1.2600 | Latest release April 21 2008. Old laptop we had for home use. It has 158 installed applications. Although Windows XP is not supported anymore by Microsoft, we tested it because most ATM machines still use it [33]. |
| Windows Server 2008 R2 | Service Pack 1 version 6.1.7601.65536. Widely used for servers. This was running on a VM as freshly installed OS with 7 reinstalled apps. |
| Windows 7 | Service Pack 1 version 6.1.7601.65536. Widely used for users. This was running on a VM as freshly installed OS with 7 reinstalled apps. |

#### 6.2.2 Accuracy Results

We performed accuracy analysis according to the previously discussed methodology in Section 6.1.1. Table 22 shows our accuracy analysis for operating system vulnerabilities. The high increase of discovered vulnerabilities by CVS is expected because it scans the whole machine's PVCs, unlike the other scanners that scan network interface PVCs only. Average accuracy for Nessus, Nexpose, and CVS was 2.16%, 2.23%, and 57.83% (respectively). Compared to Nessus, CVS increased accuracy by an average of 2620%. Compared to Nexpose, CVS increased accuracy by an average of 2720%. Many applications do not expose their services to network interfaces but are still vulnerable. This gives a false sense of security to users. Using traditional scanners gives such impression while CVS does not.

## 7 SECURITY ANALYSIS

In Section 3, we discussed our threat model and security goals. Communication between CVS-server and CVS-client uses the latest Transport Layer Security (TLS) specs. CVS can also use symmetric keys instead of TLS, but the default implementation uses TLS. Network communication is encrypted using the Advance Encryption Standard - Galois Counter Mode AES-GCM [34]. AES-GCM works as an

TABLE 21: Performance comparison between CVS, Nessus, and Nexpose (Insight VM)

| Device | Try | Vulnerability Scanner Name | | | | | | CVS Improvements | | | |
|---|---|---|---|---|---|---|---|---|---|---|---|
| | | Nessus 7.1.3 (#120) | | Nexpose 6.5.3.0 | | CVS 1.1.0. | | △ Over Nessus | | △ Over Nexpose | |
| | | Time | Pckts | Time | Pckts | Time | Pckts | Time% | Pckts% | Time% | Pckts% |
| Windows XP SP3 2600 | 1st | 1152 | 156175 | 276 | 144252 | 193 | 54 | ↓83.25 | ↓99.97 | ↓30.07 | ↓99.96 |
| | 2st | 1224 | 166837 | 288 | 148584 | 3 | 50 | ↓99.75 | ↓99.97 | ↓98.96 | ↓99.97 |
| Windows 7 SP1 | 1st | 962 | 136048 | 156 | 145452 | 107 | 44 | ↓88.88 | ↓99.97 | ↓31.41 | ↓99.97 |
| | 2st | 852 | 136008 | 164 | 146278 | 3 | 36 | ↓99.65 | ↓99.97 | ↓98.17 | ↓99.98 |
| Windows Server 2008 R2 SP1 | 1st | 1153 | 149750 | 166 | 141386 | 111 | 40 | ↓90.37 | ↓99.97 | ↓33.13 | ↓99.97 |
| | 2st | 1113 | 149950 | 181 | 146242 | 4 | 36 | ↓99.64 | ↓99.98 | ↓97.79 | ↓99.98 |
| | | | | | | | | $\overline{\Delta}$: ↓93.59% | ↓99.97% | ↓64.92% | ↓99.97% |

Time is in Seconds. Pckts: network traffic packets.

TABLE 22: Accuracy comparison between CVS, Nessus, and Nexpose (Insight VM)

| Device | Discovered Vulnerabilities by | | | | Accuracy (%) | | | CVS Improvement | | Comment |
|---|---|---|---|---|---|---|---|---|---|---|
| | Nessus | Nexpose | CVS | Actual* | Nessus | Nexpose | CVS | Nessus(%) | Nexpose(%) | |
| Windows XP SP3 2600 | 9 | 6 | 261 | 391 | ≤2.3 | ≤1.5 | 60 | ↑2500 | ↑3900 | CVS had 231/261 matches |
| Windows Server 2008 R2 SP1 | 4 | 5 | 95 | 180 | ≤2.2 | ≤2.7 | 26.6 | ↑1109 | ↑885 | CVS had 48/180 matches |
| Windows 7 SP1 | 4 | 5 | 173 | 199 | ≤2 | ≤2.5 | 86.9 | ↑4245 | ↑3376 | CVS had 173/199 matches |
| | | | | $Accuracy$: | 2.16 | 2.23 | 57.83 | ↑2620 | ↑2720 | |

*Based on [32]

authenticated mode of operation that has internal integrity checks. AES-GCM is widely used and recommended by IETF RFCs for IPsec, SSH, and TLS, and it is provably secure [35, 36]. Table 23 shows our encryption setup parameters.

In CVS, we assume the VSS host is secure against internal threats, and system administrators are considered honest. Addressing semi-honest or malicious VSS will be considered in future work. Communication between VSC and VSS encompasses two main phases: scan job request, and scan result request. Figure 9 shows communication messages between VSC and VSS.

First, the VSC requests a scan job from the VSS by sending two versions of its identifier: $VSC_{ID}^a$ and $VSC_{ID}^b$. $VSC_{ID}^a$ is sent to allow the server to identify the shared key directly. (In implementation it will be encrypted with a session key derived from the VSS public keys (e.g., RSA)). The primary need for the shared key is authentication (e.g., it can be a user hashed and salted password). $VSC_{ID}^b$ is used to verify that adversaries did not change $VSC_{ID}^a$ during transit. The VSC also includes a sequence number (SN) and a time stamp (TS) to ensure message freshness and protection from replay attacks. Additionally, the VSC includes the required scanning details ($\mathcal{RSD}$) that do not contain PII; it is used by the VSS to perform scanning. Once the VSS receives the scanning request, it will follow the scanning process shown in Figure 8. If the VSS firewall accepts the job, then the VSS will send back a token to the VSC. Otherwise, the VSC will be provided with the reject reason. The VSC can discover MITM attacks (assuming Eve changed $VSC_{ID}^a$ in message (1)) by checking the value of $VSC_{ID}^a$ in message (2); if it is different from the value in message (1), then it has been changed during transit.

Second, assuming the first phase succeeded and the VSC received the token, it will then request the scan result every specific period (e.g., 5 minutes). The VSS checks the token and validates the request. If valid, then the VSS will send back the scan results. If the VSS did not finish scanning that request, the VSS will send back a message indicating it is not ready yet. The VSS also allows a limited number of scan result requests to avoid denial-of-service attacks. We provide security analysis for the two main parties of CVS: the VSC (Section 7.1) and the VSS (Section 7.2).

### 7.1 VSC Security Analysis

Communication confidentiality between a VSC and a VSS is protected from Eve by using TLS. A VSC is protected from a replay attack from Eve by having a sequence number in the encrypted payload that depends on the previous message sent by the VSC. Message freshness is protected using a timestamp to prevent Eve from exploiting previously recorded messages. We assume both the VSC and VSS are time synchronized (including different time zones) with a specific allowance window $\delta_t$ (e.g., 1 minute). Communication integrity between a VSC and a VSS is protected through Galois Message Authentication Code (GMAC) used in the Advance Encryption Standard - Galois Counter Mode (AES-GCM). A VSC can discover a Main-in-the-Middle (MITM) attacks by comparing the $VSC_{ID}^a$ from message (1) and message (2). (The server will always detect it as impersonation attack and reject it.)

### 7.2 VSS Security Analysis

Communication confidentiality between a VSS and a VSC is protected from adversaries using TLS. Each VSC shares a unique symmetric key with the VSS for authentication. The VSS is protected from impersonation attacks by checking the VSC identifier $VSC_{ID}^a$ and $VSC_{ID}^b$. If they are not equal, then an adversary is launching an impersonation attack. The VSS is protected from replay attacks through checking the message freshness, using the time stamp and the sequence number. If a device is sending many valid requests, it will be blocked for an exponentially increasing period. Communication integrity is protected using the GMAC in AES-GCM.

TABLE 23: Parameters used for AES-GCM and GMAC

| Parameter | Key Length | Remarks |
| --- | --- | --- |
| Encryption Key Size | 128 bits | Sufficient until the year 2076. [35, 37, 38]. |
| GMAC Output Size | 128 bits | Sufficient [35]. |
| Nonce/IV Size | 128 bits | As per GCM algorithm (see [34]). |
| Salt Size | 128 bits | Mixed with the password. Increase protection against dictionary attacks. |
| Iterations | 100 | Number of iterations for the mixing function. |

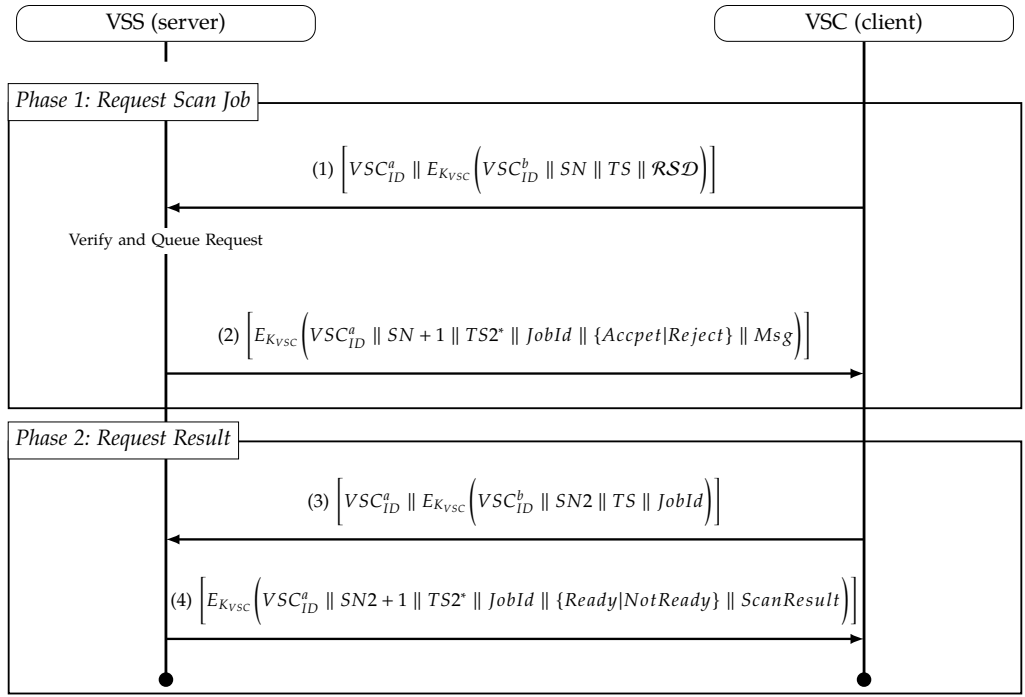

Fig. 9: Communication diagram between a VSC and a VSS in CVS. SN: Sequence number. TS: Timestamp. RSD: Required scanning details.

## 8 CONCLUSIONS

Traditional vulnerability scanning methods are time-consuming and indecisive, and negatively affect network performance by generating high network traffic. In this paper, we presented a novel vulnerability scanner called Calcium Vulnerability Scanner (CVS) that is time-efficient, simple, accurate, and safe. Our contributions to vulnerability scanning are the following: (i) minimize its required time and network traffic (compared to current technologies, we reduced the former by an average of 79% and the latter by 99.9%), (ii) increase its accuracy (compared to current technologies, we improved it by an average of 2600%), and (iii) enable the scanner to learn from previous scans in order to reduce future scanning time and enhance accuracy (compared to current technologies (that did not support learning), CVS reduced scanning time by an average of 97% through learning). CVS should enhance and facilitate vulnerability scanning and enable the implementing of scanning on a large scale efficiently. We envision that this remodeling will lead to faster, more scalable, and easier deployment of scanners for large scale big data servers, and cloud computing. Security standards can be revisited again to adjust the scanning frequency requirements which can be increased using our proposed approach to provide a better security measures. CERTs will be able to conduct more frequent scanning on larger scales and hence yield a better response time.

## ACKNOWLEDGEMENTS

An earlier version of CVS was called Push-Based Vulnerability Scanner (PBVS) and was presented in Sari Sultan's master's thesis at Kuwait University [39].